\documentclass{article}
\usepackage{frascatiphys}
\usepackage{graphicx}
\usepackage{amsmath}
\usepackage{multirow}
\usepackage{amssymb}
\begin{document}
\title{ 
PROSPECTS FOR DOUBLE HIGGS PRODUCTION
}
\author{
Giuliano Panico\\
{\em IFAE, Universitat Aut\` onoma de Barcelona,}
{\em Bellaterra, Barcelona, Spain}
}
\maketitle
\baselineskip=11.6pt
\begin{abstract}
A concise review of the double Higgs production channel at the LHC and at future hadron and lepton
machines is presented.
\end{abstract}
\baselineskip=14pt

\section{Introduction}

Double Higgs production is one example of scattering process that can disclose key information
on the electroweak symmetry breaking dynamics, in particular its underlying symmetries and strength.
It is one of the few channels that can give direct access to the quartic couplings
among two Higgs bosons and a pair of gauge bosons or of top quarks, as well as to the Higgs trilinear
self-coupling.

Due to the small cross section, the precision achievable at the LHC on these couplings
is quite limited. The large increase in cross section at high-energy hadron machines and the
improved precision possible at future lepton colliders could overcome the LHC limitations
providing an ideal environment to test this process.

In the absence of light new states, the new-physics effects can be pa\-ra\-me\-tri\-zed via
low-energy effective Lagrangians. Two formulations are useful for the study of
Higgs physics\cite{Contino:2013kra}.
The first one, the ``linear'' Lagrangian, is based on the assumption that the Higgs is part of an
$\mathrm{SU}(2)_L$ doublet, as in the SM.
In the second, more general formulation, $\mathrm{SU}(2)_L \times \mathrm{U}(1)_Y$ is
non-linearly realized, hence the name of ``non-linear'' Lagrangian, and the physical Higgs is a singlet
of the custodial symmetry, not necessarily part of a weak doublet.
The run 1 LHC indicates that the couplings of the newly discovered boson are close
to the values predicted for the SM Higgs. This clearly motivates the use of the linear Lagrangian for
future studies.
Indeed, small deviations from the SM are naturally expected if the Higgs boson belongs to a doublet,
provided the new states are much heavier than the weak scale.
The non-linear formulation is still useful, however, when large deviations 
in the Higgs couplings are allowed. This is especially true for double Higgs production,
from which additional couplings not accessible via single Higgs processes can be
extracted\cite{HH_CH,Azatov:2015oxa}.

In the linear Lagrangian, the operators can be organized as
\begin{equation}
{\cal L}_{\textrm{lin}} = {\cal L}_{\textrm{SM}} + \Delta {\cal L}_6 + \Delta {\cal L}_8 + \dots
\end{equation}
The lowest-order terms coincide with the usual SM Lagrangian ${\cal L}_{\textrm{SM}}$,
whereas ${\cal L}_{n}$ contains the deformations due to operators of dimension $n$, with $n > 4$.
For our purposes it is sufficient to focus on the operators involving the Higgs boson.
The ones in $\Delta {\cal L}_6$ relevant for double Higgs production are (for simplicity we
only include the CP-conserving operators)
\arraycolsep = 2pt
\begin{eqnarray}
\Delta {\mathcal L}_6 &\supset & \frac{\overline c_H}{2 v^2} \left[\partial_\mu (H^\dagger H)\right]^2
+ \frac{\overline c_u}{v^2} y_u H^\dagger H \overline q_L H^c u_R
- \frac{\overline c_6}{v^2} \frac{m_h^2}{2 v^2} (H^\dagger H)^3\nonumber\\
&& +\; \frac{c_g}{m_W^2} g_s^2 H^\dagger H G_{\mu\nu}^a G^{a\; \mu\nu}\,,
\label{eq:linear_lagr}
\end{eqnarray}
where $H$ denotes the Higgs doublet, $v = 246\ \mathrm{GeV}$ and $m_h = 125\ \mathrm{GeV}$
is the Higgs mass. The linear Lagrangian relies on a double expansion. The first one is an expansion
in derivatives, in which higher-order terms are suppressed by additional powers of $E^2/m_*^2$.
To derive this estimate we assumed that the new dynamics can be broadly
characterized by a single mass scale $m_*$, at which new states appear, and by one coupling
strength $g_*$ (this is the so called SILH power counting\cite{Giudice:2007fh}). The second
expansion is in powers of the Higgs doublet: each extra insertion is
weighted by a factor $1/f \equiv g_*/m_*$. In order to be under control, the linear Lagrangian requires
$E^2/m_*^2 < 1$ and $v/f < 1$.

In the case of the non-linear Lagrangian, the relevant operators are
\begin{eqnarray}
{\mathcal L} & \supset & \left(m_W^2 W_\mu W^\mu + \frac{m_Z^2}{2} Z_\mu Z^\mu\right)
\left(1 + 2 c_V \frac{h}{v} + c_{2V} \frac{h^2}{v^2}\right)
- c_3 \frac{m_h^2}{2v}h^3\nonumber\\
&& -\; m_t \overline t t \left( 1 + c_t \frac{h}{v} + c_{2t} \frac{h^2}{2 v^2}\right)
+ \frac{g_s^2}{4 \pi^2}\left(c_g \frac{h}{v} + c_{2g} \frac{h^2}{2v^2}\right) G_{\mu\nu}^a G^{a\; \mu\nu}\,,\label{eq:non-linear_lagr}
\end{eqnarray}
where $h$ denotes the physical Higgs field (with vanishing expectation value).
With respect to the linear parametrization, the operators in Eq.~(\ref{eq:non-linear_lagr})
effectively resum all the corrections of order $v^2/f^2$. The non-linear Lagrangian only relies
on the derivative expansion, but not on the expansion in powers of the Higgs field.
When the linear and non-linear parametrizations are both valid, the coefficients of the two
effective Lagrangians are related by
\begin{eqnarray}
& \displaystyle c_t = 1 - \overline c_H/2 - \overline c_u\,,
\qquad
c_{2t} = -(\overline c_H + 3\, \overline c_u)/2\,,
\qquad
c_3 = 1 - 3\,\overline c_H/2 + \overline c_6\,,\nonumber\\
& \displaystyle  c_g = c_{2g} = \overline c_g \left({16\pi^2}/{g^2}\right)\,,
\qquad
c_V = 1 - {\overline c_H}/{2}\,,
\qquad
c_{2V} = 1 - 2\, \overline c_H\,.
\end{eqnarray}
Notice that single operators in the linear Lagrangian induce correlated modifications in
different Higgs vertices. For instance the ${\mathcal O}_u$ operator, which gives
a modification of the top Yukawa, also generates a new quartic interaction $\overline t t hh$.

\section{Double Higgs at hadron colliders}

Double Higgs production at hadron colliders is mainly due to three processes:
Gluon Fusion (GF), Vector Boson Fusion (VBF) and $tthh$ associated production.
In the following we will focus on the GF and VBF channels, for which dedicated analyses
at high-energy colliders exist. The $tthh$ channel,
for which only LHC studies are currently available\cite{Englert:2014uqa},
can provide some information on the Higgs trilinear coupling,
but it seems not competitive with the GF
channel.

\subsection{Gluon fusion}

The GF channel is the dominant production mode at hadron colliders.
The NNLO SM cross section at the $14\ \mathrm{TeV}$ LHC is
$\sigma_{\textrm{SM}} \simeq 37\,\mathrm{fb}$,
while it becomes $\sigma_{\textrm{SM}} \simeq 1.5\,\mathrm{pb}$ at a $100\ \mathrm{TeV}$ collider.
The relatively small cross sections imply that only a few final states are relevant.
In spite of the small branching fraction ($BR \simeq 0.264\%$) the
$hh \rightarrow \gamma\gamma b\overline b$ channel has been recognized as the most promising
one due to the clean signal and small backgrounds\cite{HH_had_bbgaga,Azatov:2015oxa}.
Other channels, whose exploitation is more difficult due to the large backgrounds, have been also
considered, among which $hh \rightarrow b\overline b \tau^+ \tau^-$,
$hh \rightarrow b\overline b W W^*$ and $hh \rightarrow b\overline b b \overline b$\cite{other_HH}.
Due to the larger cross section these channels could be relevant for an analysis of the high-energy tail of the
kinematic distributions, where boosted jet techniques could enhance the signal reconstruction efficiency.

The GF channel is sensitive to several new-physics effects.
In the non-linear formalism, it depends on the Higgs self-coupling ($c_3$),
on the top couplings ($c_t$, $c_{2t}$) and on the contact interactions with the
gluons ($c_g$, $c_{2g}$). It is thus a privileged channel to test the non-linear
Higgs couplings ($c_3$, $c_{2t}$, $c_{2g}$) that can not be directly accessed in
single-Higgs processes.
Interestingly, the various new physics effects affect in different ways the kinematic distributions
(in particular, the Higgs-pair invariant mass $m_{hh}$).
An exclusive analysis taking into account the $m_{hh}$ distribution can thus be used to disentangle
the various coefficients in the effective Lagrangian\cite{Azatov:2015oxa}.
This is relevant at high-energy colliders, where the sizable cross section
allows to reconstruct the $m_{hh}$ distribution, it is instead of limited applicability at the LHC
due to the small number of signal events.

\begin{table}
\centering
\begin{tabular}{ccccc}
\rule[-.4em]{0pt}{1.em}& LHC$_{14}$ & HL-LHC & FCC$_{100}$ & Reference\\
\hline
\hline
\rule[-.6em]{0pt}{1.7em}$\overline c_6$ & $[-1.2, 6.1]$ & $[-1.0, 1.8] \cup [3.5, 5.1]$ & {$[-0.33,0.29]$} &
{\scriptsize Azatov {\it et al.}\cite{Azatov:2015oxa}}\\
\hline
\rule{0pt}{1.15em}$\Delta c_{2V}$ & $[-0.18, 0.22]$ & $[-0.08, 0.12]$ & $[-0.01, 0.03]$ &
{\scriptsize Contino {\it et al.}\cite{Contino_talk}}\!\\
\end{tabular}
\caption{\it Estimated precision on the Higgs trilinear coupling $\overline c_6$ and
$\Delta c_{2V} = c_{2V} -1$ at hadron machines. The table reports the $68\%$ probability intervals.}
\label{tab:perc_had}
\end{table}

To conclude the discussion we report in table~\ref{tab:perc_had}
the precision on the determination of the Higgs trilinear coupling $\overline c_6$
for three benchmark scenarios: $14\ \mathrm{TeV}$ LHC
with $L = 300\,\mathrm{fb}^{-1}$ integrated luminosity (LHC$_{14}$), high-luminosity LHC
with $L = 3\,\mathrm{ab}^{-1}$ (HL-LHC) and a future $100\ \mathrm{TeV}$ $pp$ collider with
$L = 3\,\mathrm{ab}^{-1}$ (FCC$_{100}$). It is important to stress that the precision
on the $\overline c_6$ coefficient is affected by the uncertainty on the other parameters
in the effective Lagrangian and
in particular on the top Yukawa, $\overline c_u$ (the result in table~\ref{tab:perc_had} was derived
by assuming $\Delta \overline c_u \simeq 0.05$). With no uncertainty on $\overline c_u$, the
Higgs trilinear coupling could be extracted at FCC$_{100}$ with precision $\Delta \overline c_6 \simeq 0.18$.

\subsection{Vector boson fusion}

The VBF channel is sensitive to the Higgs self-coupling $c_3$ and, more importantly,
to the single and double Higgs coupling to the vector bosons ($c_V$, $c_{2V}$).
Analogously to $WW$ scattering, a modification of the Higgs coupling to the gauge fields
spoils the cancellation present in the SM, so that the VBF amplitude
grows at high energy as ${\mathcal A} \sim {\hat s}/{v^2} (c_V^2 - c_{2V})$. The tail of the distribution
is thus particularly sensitive on $c_V$ and $c_{2V}$. The Higgs trilinear, on the contrary,
affects the $m_{hh}$ distribution mostly at threshold and has a limited impact.

The small cross section forces to consider Higgs decay channels with large branching fractions.
The most relevant final state is $hh \rightarrow 4b$. Estimates of the precision achievable on
$c_{2V}$ are given in table~\ref{tab:perc_had} for three benchmark scenarios.

\section{Double Higgs at lepton colliders}

The main channels for double Higgs production at lepton colliders are Double Higgs-Strahlung (DHS)
and Vector Boson Fusion (VBF).
The DHS channel is dominant for center of mass energies below $s \lesssim 1\ \mathrm{TeV}$,
while above this threshold the VBF cross section becomes the largest one\cite{Tian:2013yda}.

\begin{table}
\centering
\begin{tabular}{c@{\hspace{1.25em}}c@{\hspace{1.25em}}l@{\hspace{1.em}}c@{\hspace{1.em}}c}
\rule[-.4em]{0pt}{.5em}& COM Energy & \hspace{.5em} Precision & Process & Reference\\
\hline
\hline
\rule{0pt}{1.65em}\multirow{3}{*}{\vspace{-1.25em} \bf{ILC}} & \parbox{.17\textwidth}{\centering $500\ \mathrm{GeV}$\\ \rule{0pt}{1em}\scriptsize $[L = 500\ \mathrm{fb}^{-1}]$}
& $\Delta c_3 \sim 104\%$ & DHS & {\scriptsize ILC TDR, Volume 2\cite{Baer:2013cma}}\\

\rule{0pt}{1.65em}& \multirow{2}{*}{\vspace{-.4em}\parbox{.16\textwidth}{\centering $1\ \mathrm{TeV}$\\ \rule{0pt}{1.em}\scriptsize $[L = 1\ \mathrm{ab}^{-1}]$}} &
$\Delta c_3 \sim 28\%$ & VBF & {\scriptsize ILC TDR, Volume 2\cite{Baer:2013cma}}\\

\rule[-.55em]{0pt}{1.65em}& & $\Delta c_{2V} \sim 20\%$ & DHS & {\scriptsize Contino {\it et al.}\cite{Contino:2013gna}}\\
\hline
\rule{0pt}{1.05em}\multirow{4}{*}{\vspace{-2.25em} \bf{CLIC}} & \multirow{2}{*}{\vspace{-.4em}\parbox{.17\textwidth}{\centering $1.4\ \mathrm{TeV}$\\ \rule{0pt}{1em}\scriptsize $[L = 1.5\ \mathrm{ab}^{-1}]$}}
& $\Delta c_3 \sim 24\%$ & \multirow{4}{*}{\vspace{-2.25em} VBF} & \multirow{4}{*}{\vspace{-2.25em} \scriptsize P. Roloff (CLICdp Coll.)\cite{Roloff}}\\

\rule{0pt}{1.1em}& & $\Delta c_{2V} \sim 7\%$ & &\\

\rule{0pt}{1.65em}& \multirow{2}{*}{\vspace{-.4em}\parbox{.17\textwidth}{\centering $3\ \mathrm{TeV}$\\ \rule{0pt}{1.em}\scriptsize $[L = 2\ \mathrm{ab}^{-1}]$}} & $\Delta c_{3} \sim 12\%$ & &\\

\rule{0pt}{1.1em}& & $\Delta c_{2V} \sim 3\%$ & &
\end{tabular}
\caption{\it Expected $68\%$ CL precision on the Higgs trilinear coupling $c_3$ and on the $c_{2V}$ coupling
at future lepton colliders.}
\label{tab:prec_ee}
\end{table}

Both production channels are sensitive to deviations in the Higgs trilinear coupling and in the
double Higgs coupling to vector bosons. The expected precision on the determination
of $\Delta c_3$ and $\Delta c_{2V}$ for different benchmark scenarios are
listed in table~\ref{tab:prec_ee}. In order to obtain a fair determination of these parameters
a center of mass energy $s \gtrsim 1\ \mathrm{TeV}$ and an integrated luminosity
$L \gtrsim 1\ \textrm{ab}^{-1}$ are necessary. With these minimal requirements a precision
of the order $20 - 30\%$ can be achieved. Further improvements in the collider energy
could significantly boost the precision on $c_{2V}$, up to a $\sim 3\%$ accuracy,
since the effects mediated by this coupling are enhanced at high $m_{hh}$.
The deviations in the Higgs trilinear coupling, on the contrary, affect mostly
the distribution at threshold, hence an improvement in the precision at higher energies
is mainly related to the luminosity increase.
The precision on $c_3$ and $c_{2V}$ that can be obtained at lepton machines with
$s \gtrsim 1\ \mathrm{TeV}$ is roughly comparable to the one estimated for
a $100\ \mathrm{TeV}$ hadron collider (see the FCC$_{100}$ column
in table~\ref{tab:perc_had}).

\end{document}